\documentclass[journal,twoside,web]{ieeecolor}
\usepackage{tmi}
\usepackage{cite}
\usepackage{amsmath,amssymb,amsfonts}
\usepackage{algorithmic}
\usepackage{graphicx}
\usepackage{textcomp}
\usepackage{threeparttable}
\usepackage{bbding}
\usepackage{float}
\usepackage{mathrsfs}
\def\BibTeX{{\rm B\kern-.05em{\sc i\kern-.025em b}\kern-.08em
    T\kern-.1667em\lower.7ex\hbox{E}\kern-.125emX}}
\begin{document}
\title{ComptoNet: An End-to-End Deep Learning Framework for Scatter Estimation in Multi-Source Stationary CT}
\author{Yingxian~Xia, Zhiqiang~Chen, Li~Zhang, Yuxiang~Xing, and Hewei~Gao
\thanks{The authors are with the Department of Engineering Physics, Tsinghua University, Beijing 100084, China, and also with the Key Laboratory of Particle and Radiation Imaging (Tsinghua University), Ministry of Education, Beijing 100084, China. (e-mails: xyx20@mails.tsinghua.edu.cn; czq@mail.tsinghua.edu.cn; zli@mail.tsinghua.edu.cn; xingyx@mail.tsinghua.edu.cn; hwgao@ tsinghua.edu.cn.)}
}

\maketitle

\begin{abstract}
Multi-source stationary computed tomography (MSS-CT) offers significant advantages in medical and industrial applications due to its gantry-less scan architecture and/or capability of simultaneous multi-source emission.
However, the lack of anti-scatter grid deployment in MSS-CT results in severe forward and/or cross scatter contamination, presenting a critical challenge that necessitates an accurate and efficient scatter correction.
In this work, ComptoNet, an innovative end-to-end deep learning framework for scatter estimation in MSS-CT, is proposed, which integrates Compton-scattering physics with deep learning techniques to address the challenges of scatter estimation effectively.
Central to ComptoNet is the Compton-map, a novel concept that captures the distribution of scatter signals outside the scan field of view, primarily consisting of large-angle Compton scatter. In ComptoNet, a reference Compton-map and/or spare detector data are used to guide the physics-driven deep estimation of scatter from simultaneous emissions by multiple sources.
Additionally, a frequency attention module is employed for enhancing the low-frequency smoothness.
Such a multi-source deep scatter estimation framework decouples the cross and forward scatter.
It reduces network complexity and ensures a consistent low-frequency signature with different photon numbers of simulations, as evidenced by mean absolute percentage errors (MAPEs) that are less than $1.26\%$.
Conducted by using data generated from Monte Carlo simulations with various phantoms, experiments demonstrate the effectiveness of ComptoNet, with significant improvements in scatter estimation accuracy (a MAPE of $0.84\%$).
After scatter correction, nearly artifact-free CT images are obtained, further validating the capability of our proposed ComptoNet in mitigating scatter-induced errors.
\end{abstract}


\section{Introduction}
\IEEEPARstart{M}{ulti}-source stationary computed tomography (MSS-CT) is an innovative CT imaging configuration under investigations for decades \cite{spronk2021evaluation, duan2020novel, cramer2018stationary,wu2017swinging,de2016multisource,wang2024iterative}, with novel architectures and potential applications keeping emerging in recent years, such as symmetric-geometry CT \cite{zhang2020stationary}, compact and non-circular stationary Head CT \cite{mcskimming2023adaptive}, dual-ring MSS-CT \cite{Chen2023Millisecond}.
MSS-CT offers a unique opportunity to significantly boost the temporal resolution of image acquisition through two key mechanisms: electronically controlled gantry-less scanning and/or simultaneous multi-source emission.
The employment of electronically controlled gantry-less scanning avoids the use of slip ring gantry technology and enables MSS-CT systems to surpass the upper bounds of mechanical rotation velocities, attaining a CT scan in just tens of milliseconds.
While, the multi-source nature of MSS-CT inherently supports simultaneous beam emission, further augmenting scanning speed.
However, simultaneous multi-source emission would face a major challenge in scatter correction, as the primary signal will be not only contaminated by scatter originated from the same source as the primary (forward scatter) but also by scatter originated from other sources (cross scatter).
As a result, to overcome the scatter contamination problem and achieve high quality CT images, it is extremely crucial to develop highly effective and efficient scatter estimation and correction techniques for MSS-CT.

In general, scatter problems in CT can be addressed through hardware-based scatter rejection or software-driven scatter estimation and correction \cite{ruhrnschopf2011general}.
Hardware-based methods minimize the impact of scatter signals by using physical accessories such as anti-scatter grids (ASG), collimators, and bowtie filters \cite{siewerdsen2001cone, kyriakou2007efficiency, rinkel2007coupling, wiegert2006scattered}.
For MSS-CT, implementation of a hardware-based method like ASG can be very challenging, due to the significant changes in the relative position of source to detector across different scan views.
In contrast, software-based methods focus on using algorithms and models to accurately estimate and subtract scatter signals from the scatter-contaminated data, thereby recovering primary signals.
Such approaches primarily falls into three categories: Monte Carlo simulations (MC), scatter kernel superposition (SKS) techniques, and deep learning-based methods.
MC simulations, in general, can be highly accurate for scatter estimation but are computationally demanding \cite{kyriakou2006combining, bootsma2015efficient, qin2023correlated}.
While known for their computational efficiency and practicality, SKS techniques can be less accurate in complex cases.  \cite{hansen1997extraction,star2009efficient, sun2010improved, kim2015data}.

Deep learning methodologies are anticipated to achieve high precision and computational efficiency for CT scatter correction.
Pioneeringly, both ScatterNet \cite{hansen2018scatternet} and Deep Scatter Estimation (DSE) method \cite{maier2018deep} have adopted a similar Unet-like architecture, which is trained to predict outputs using the acquired projection data as inputs. The effectiveness of Unet architecture has been further substantiated under different cases \cite{maul2022learning, maier2019real}. However, despite the promise shown by Unet-based scatter removal techniques, their lack of physics constraints may lead to spurious outcomes and high-frequency artifacts, indicating a need for further refinement in these methods.

Mechanisms that can impose constraints on the Unet architecture have been explored recently.
B-splines are integrated into neural networks, ensuring smoother and more reliable predictions \cite{roser2021x}.
A novel scatter correction algorithm has been proposed, combining Convolutional Neural Networks (CNNs) with the Swin Transformer \cite{zhang2023image}.
Another method introduces the generation of scatter amplitude and width maps from projection images using a neural network, with the final scatter map computed through a convolution process \cite{zhuo2023scatter}.
Physics-inspired deep learning methods have also been developed for scatter estimation and correction, drawing on the underlying physics of X-ray scattering for data processing pipelines, network architectures, and loss function designs \cite{iskender2022scatter}.

For Multi-source CT, addressing both forward and cross scatter correction presents greater challenges than regular cases.
Research has shown that dual-source CT, for instance, can cause cross scatter artifacts that are almost twice as pronounced as those in single-source CT \cite{engel2008x}.
Some studies suggest that cross-scatter is mainly generated near the object surface and can be reduced through model-based scatter correction methods \cite{petersilka2010strategies}.
A beam-stopper-array-based method has been implemented for simultaneous online correction of forward and cross scatter in multi-source CT \cite{gong2017x}. Additionally, a physics model-based and iterative framework has been presented for calculating X-ray scatter signals in both forward and cross directions \cite{gong2017physics}.
Furthermore, an adaptive kernel strategy has been introduced for efficient estimation of large, non-isotropic X-ray scatter in stationary multi-source CT \cite{mcskimming2023adaptive}. 
More recently, a Unet-based method has been employed for simultaneous forward and cross-scatter correction \cite{erath2021deep}, but unfortunately, physical differences between forward and cross scatter was not taken into account.

In this study, to achieve accurate deep scatter estimation for dual-ring MSS-CT, a novel end-to-end deep learning framework, ComptoNet, is proposed, which integrates physical prior knowledge into the model.
ComptoNet consists of two networks sequentially: a Conditional Encoder-Decoder Network (CED-Net) for cross scatter estimation and a frequency Unet for forward scatter estimation.
The physical prior knowledge in ComptoNet manifests as Compton-map, which is mainly composed of large-angle Compton scatter signals.
In the deep cross scatter estimation process, reference Compton-map and/or spare detector data are employed for realizing a guidance mapping decoder.
Frequency attention module used in the deep forward scatter estimation preserves the high-frequency scatter information, contrary to conventional down-sampling methods, which leads to improved predictions.

\section{Materials and methods}

\subsection{Multi-Source Stationary CT with A Dual-Ring Shaped Architecture}

\begin{figure}[htb]
	\centering
	\includegraphics[width=9cm]{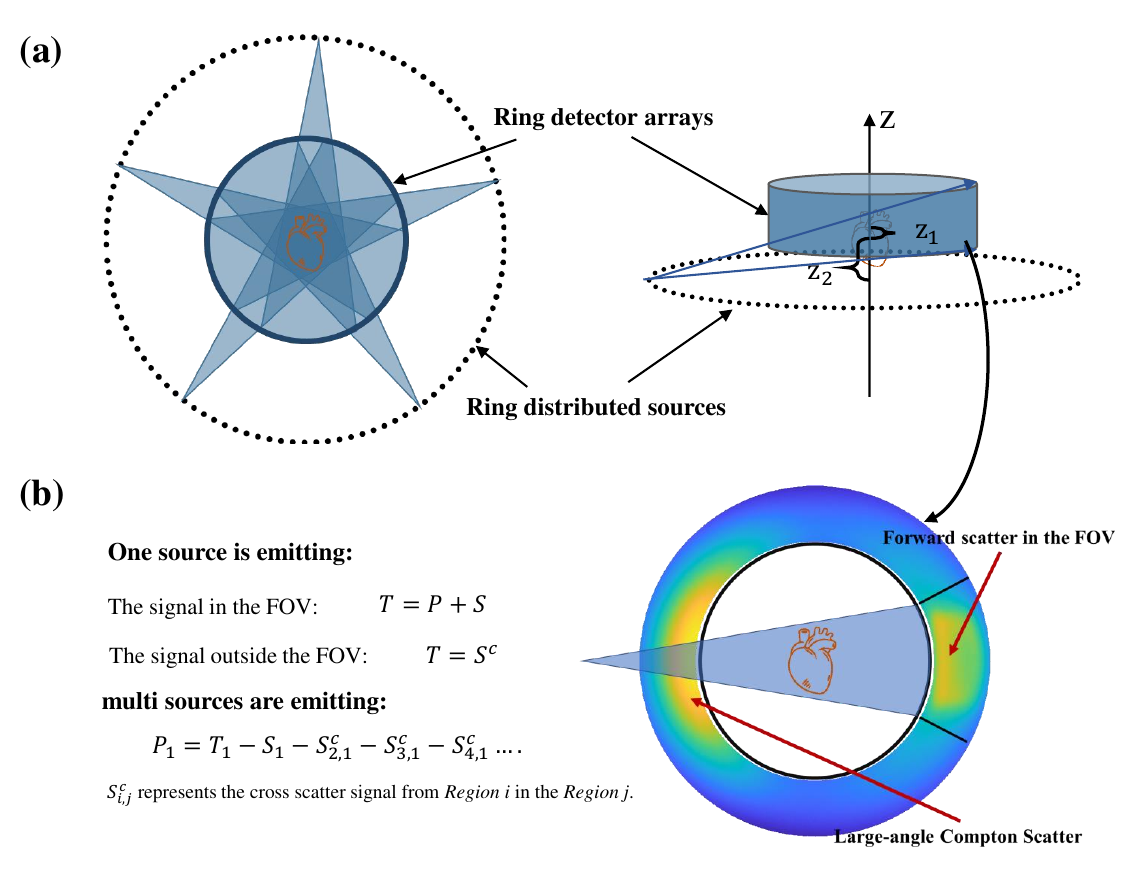}
	\caption{(a) An architecture of dual-ring shaped MSS-CT, where the ring distributed sources and ring detector arrays are arranged on two different planes for unobstructed 360-degree CT scanning. (b) The scatter distribution of dual ring MSS-CT.}
	\label{fig_DRCT} 
\end{figure}

Among various MSS-CT imaging architectures, dual-ring MSS-CT system is a prototypical and novel design, utilizing a distributed circular X-ray source and a circular detector array to acquire projections from 360-degree view as shown in Fig. \ref{fig_DRCT}(a).
It has been studied by many researchers \cite{chen2014stationary, xia2023generalized, Chen2023Millisecond, zha2024bilateral} and offering distinct advantages over other stationary CT configurations, particularly for cardiac imaging.

When the $j$-th X-ray source is emitting in MSS-CT, the total photon signal $ T^i_j $ received by the detector element $i$ can be expressed as
\begin{equation}
T^i_j = P^i_j + S^i_j,
\end{equation}
where
$P^i_j$ is the primary signal and
$S^i_j$ is the scatter signal.
For the detector element outside of the $j$-th FOV, $P^i_j = 0$.

When X-ray sources $ 1, 2, \ldots, n $ are emitting simultaneously, the total signal $ T^i $ received by the detector element $i$ can be expressed as:
\begin{equation}
	T^i = \sum_{j=1}^{n} T^i_j = \sum_{j=1}^{n} P^i_j + \sum_{j=1}^{n} S^i_j
\end{equation}

In dual-ring MSS-CT, when multiple X-ray sources are emitting, there are multiple fields of view (FOVs).
To avoid multiplexing, multiple FOVs are evenly distributed across the circular detector array.
As a result, there are two kinds of detector elements: the elements in the FOVs and the spare elements.
For the elements in the FOVs,
\begin{equation}
T^{i_m} = P^{i_m} + \sum_{j=1}^{n} S^{i_m}_j.
\end{equation}
For the spare elements,
\begin{equation}
T^{i_0} = \sum_{j=1}^{n} S^{i_0}_j.
\end{equation}
where
$i_m$ represents the element in the $m$-th FOV and
$i_0$ represents the spare element.

The primary signal $ P^{i_m}$ in the $m$-th FOV can be expressed as:
\begin{equation}
	P^{i_m} = T^{i_m} - \sum_{j=1}^{n} S^{i_m}_j.
\end{equation}

$\{S^{i_m}_j\}$ include the forward scatter signals originated from the same source (${i_m}=j$) and multiple cross scatter signals originated from other sources (${i_m} \neq j$).
The deep learning framework $\mathscr{F}$ should realize:
\begin{equation}
	\{S^{i_m}_j\} = \mathscr{F}(T^{i_m})
\end{equation}
It is clear that forward scatter $\{S^{i_m}_{i_m}\}$ is similar to $T^{i_m}$, so Unet-based method could be used to predict $S^{i_m}_{i_m}$ from $T^{i_m}$ directly.
Cross scatter signals $\{S^{i_m}_j\}_{j \neq i_m}$ have different distribution from $S^{i_m}_{i_m}$, necessitating additional analysis and design considerations.

\subsection{Compton-map}
The low-frequency characteristics of forward scatter have been consistently identified \cite{love1987scatter}.
Compared with forward scatter, cross scatter exhibit simpler distributions and is primarily surface scatter. The outer dimensions and shape of an object significantly influence cross-scatter intensity distribution \cite{kyriakou2007intensity, petersilka2010strategies}.

\begin{figure}[htb]
	\centering
	\includegraphics[width=8.5cm]{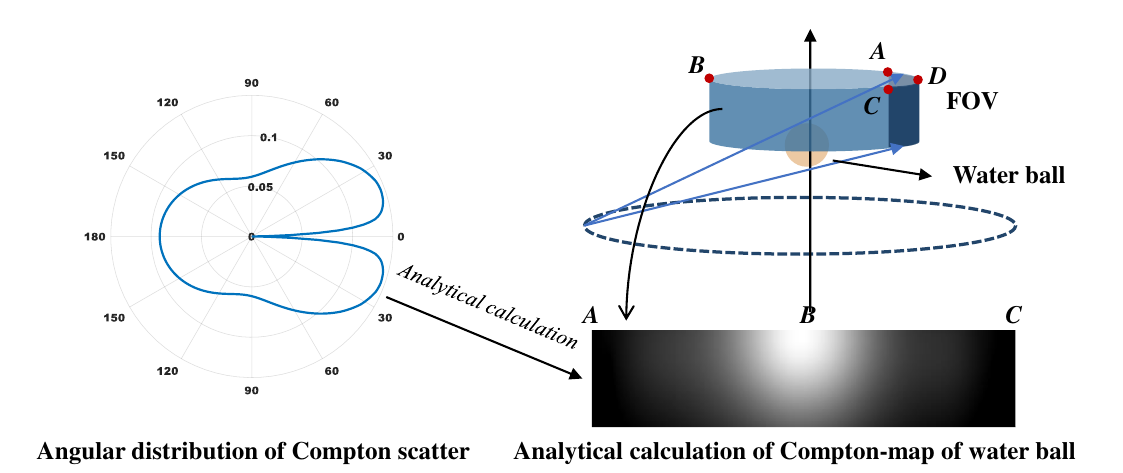}
	\caption{Compton-map of water ball from analytical physical calculation.}
	\label{fig_ComptonMap} 
\end{figure}

Considering the primary component of cross scatter, the magnitude and distribution of $\{S^{i}_j\}_{i \neq j_m}$ are denoted as Compton-maps.
For example, when one source is emitting in dual-source CBCT, the received scatter signal of the other flat panel detector that is not aligned with the emitting X-ray source is the Compton-map of dual-source CBCT.
For dual-ring MSS-CT, the received scatter signals outside the FOV of the emitting source is the Compton-map, as shown in Fig. \ref{fig_ComptonMap}.

Given that first-order large-angle Compton scattering constitutes the predominant component of Compton-maps, direct physical analytical calculations of the distribution of large-angle Compton scattering signals can be undertaken to elucidate the properties of Compton-maps.
Specifically, for large-angle Compton scatter in MSS-CT, we could integrate the large-angle Compton scatter signals originated from all voxels when the $j$-th X-ray source is emitting \cite{yao2009analytical}:
\begin{equation}
	\begin{split}
		S^i_j = \sum_{k} E_{s}
		& \left( 
		Q_{k} \frac{\mathrm{d} \sigma_{C}(\theta,E,Z)}{\mathrm{d} \Omega} \int \rho_{v} \mathrm{d} v
		\frac{A \cos \alpha}{{l_s}^2}
		\right) \cdot \\
		& e^{- \int u_s(E_{s}) dt}
		(1 - e^{-u_{det}(E_{s})d})
	\end{split}
	\label{eq1}
\end{equation}
where $Q_{k}$ is the number of primary photons at the voxel $k$, $\frac{A \cos \alpha}{{l_s}^2}$ is the solid angle subtended by the $i$ detector element, $\int \rho_{v} \mathrm{d} v$ is the line integral of the molecular density, $E_{s}$ is the energy of the scattered photon
\begin{equation}
	E_{s} = \frac{E_0}{1+\frac{E_0}{511 \mathrm{keV}}(1 - \cos \beta)},
\end{equation}
$\frac{\mathrm{d} \sigma_{C}(\theta,E,Z)}{\mathrm{d} \Omega}$ is the scattering cross section per atom
\begin{equation}
	\begin{split}
		& \frac{\mathrm{d} \sigma_{C}(\theta,E,Z)}{\mathrm{d} \Omega}
		= \frac{r_{e}^2}{2}{[1+k(1 - \cos \theta)]}^{-2} \\
		& {[1 + \cos^2 \theta + \frac{k^2 {(1 - \cos \theta)}^2}{1+k(1 - \cos \theta)}]}
		S(\frac{E}{hc} \sin \frac{\theta}{2}, Z).
	\end{split}
\end{equation}
Klein-Nishina formula modified by the incoherent scattering function was utilized to calculate Compton scatter cross section per atom.

According to Equation (\ref{eq1}), the calculation of Compton-maps $\{S^{i}_j\}_{i \neq j_m}$ for complex phantoms is challenging due to the various factors, including the scanned object, CT geometry, and physical characteristics of the source and detector.
In contrast, the calculation of the Compton-map for a water ball phantom is relatively straightforward for dual-ring MSS-CT, as demonstrated in Fig. \ref{fig_ComptonMap}.

\begin{figure}[htb]
	\centering
	\includegraphics[width=9cm]{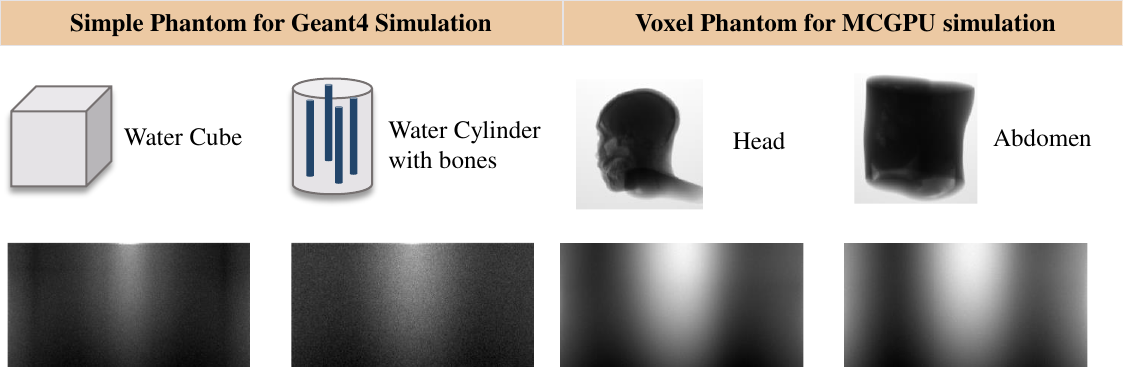}
	\caption{Compton-maps of different phantoms with Geant4 and MCGPU for the dual-ring MSS-CT geometry. They share similarity in distribution, composed of lower frequency components, and have the maximum scatter intensity observed in the upper region opposite to the FOV.}
	\label{fig_G4result} 
\end{figure}

Besides analytical physical calculation, we also conducted Monte Carlo simulations on different phantoms with Geant4 and MCGPU for the dual-ring MSS-CT geometry to investigate the properties of Compton-maps shown in Fig. \ref{fig_G4result}.

Compton-maps of MSS-CT in Fig. \ref{fig_ComptonMap} and \ref{fig_G4result} demonstrate good consistency and particularly evident in the maximum scatter intensity observed in the upper region opposite to the FOV, which indicates that the Compton-map does not have a strong correlation with the scanned object but is more closely related to the CT geometry, which is in agreement with previous analyses of the properties of cross scatter.
Therefore, we could model the Compton-map of MSS-CT as a function of CT geometry and other factors:
\begin{equation}
\{S^{i}_j\}_{i \neq j_m}  = \mathcal{R}(S_0;z),
\end{equation}
where $S_0$ is the reference Compton-map of MSS-CT, which is the Compton-map of water ball from analytical physical calculation and $z$ represents the scanned object and other physical factors.
The reference Compton-map already includes the geometric information of MSS-CT.

\begin{figure*}[htb]
	\centering
	\includegraphics[width = 14cm]{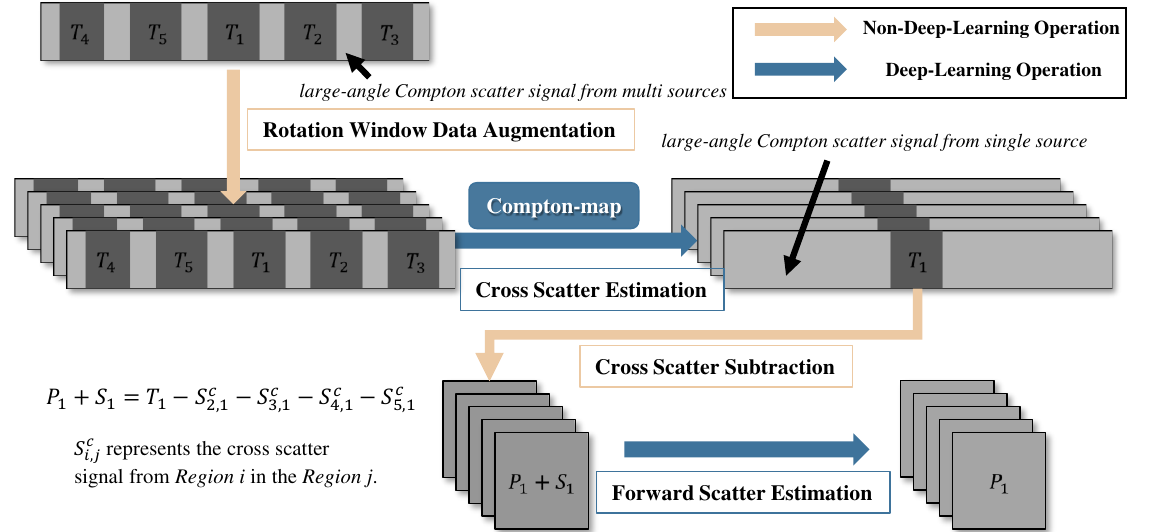}
	\caption{ComptoNet: an end-to-end deep learning framework for scatter estimation in dual-ring MSS-CT. If five sources are emitting simultaneously in the dual-ring MSS-CT, there will be five FOVs in the raw circular multi-source global intensity image. First, the global images could be rotated and divided with different FOV placed in the center for realizing data augmentation. And then, the five global images can be used to generate five corresponding cross scatter signals, under the guidance of the reference Compton-map and/or spare detectors. In the Cross Scatter Subtraction procedure, the cross-scatter signals from sources 2, 3, 4, and 5 are subtracted from the FOV 1. The rest FOVs can be done in the same manner. Finally, the primary signals are generated by subtracting forward scatter estimated by the Frequency Unet (Freq-Unet).}
	\label{fig_ComptoNet} 
\end{figure*}

\subsection{ComptoNet}

\subsubsection{The Concept of ComptoNet}
To utilize the properties of Compton-map analyzed previously and the signals from spare detectors in dual-ring MSS-CT, a deep scatter estimation framework named ComptoNet is proposed, which distinguishes and estimates the two types of scatter separately, and capitalizes on the property of similar distributions characteristic of Compton-maps.

The workflow of ComptoNet is shown in Fig. \ref{fig_ComptoNet}. For example, when there are five sources emitting simultaneously in the dual-ring MSS-CT, there will be five FOVs in the raw circular multi-source global intensity image. First, the global images could be rotated and divided with different FOV placed in the center for realizing data augmentation. And then, the five global images can be used to generate five corresponding cross scatter signals, under the guidance of the reference Compton-map and/or spare detectors.
In the Cross Scatter Subtraction procedure, the cross-scatter signals from sources 2, 3, 4, and 5 are subtracted from the FOV 1.
The rest FOVs can be done in the same manner. Finally, the primary signals are generated by subtracting forward scatter estimated by the Frequency Unet (Freq-Unet).

\subsubsection{Conditional Encoder-Decoder-Net}
\begin{figure}[htb]
	\centering
	\includegraphics[width = 9cm]{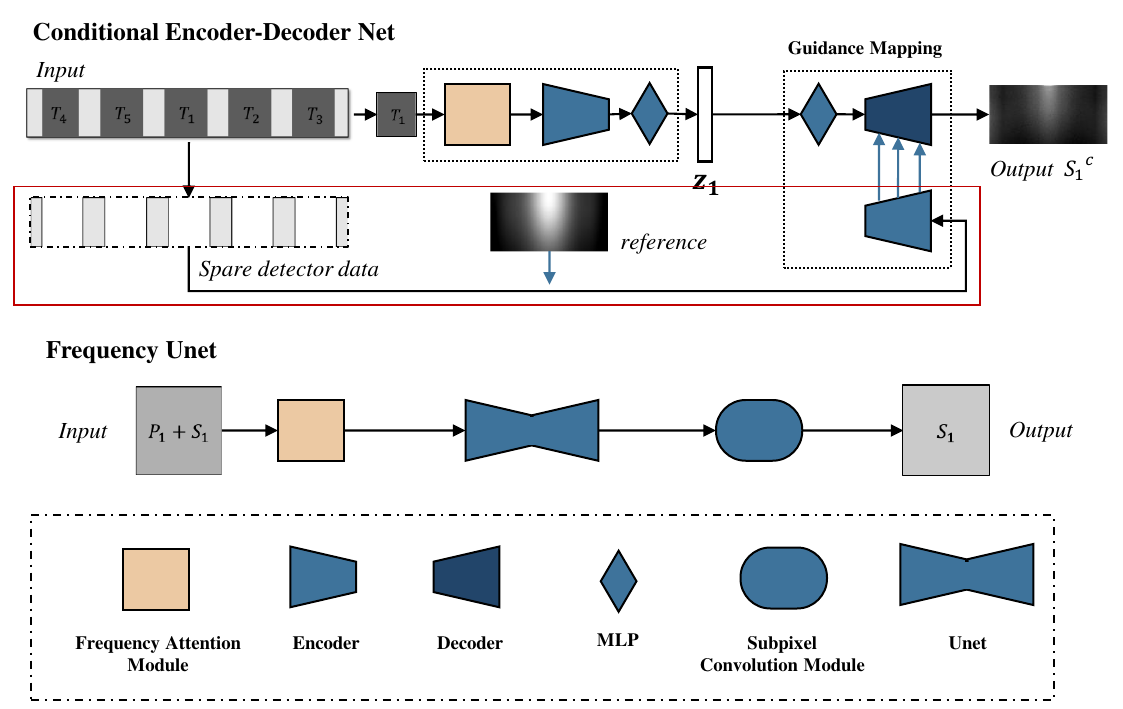}
	\caption{The architecture of Conditional Encoder-Decoder Net (CED-Net) for cross scatter estimation and frequency Unet for forward scatter estimation. In CED-Net, the decoder accordingly—using reference Compton-map (large-angle Compton scatter signal of water sphere) and spare detector data as input (red square), encoding it, and concatenating features of different dimensions into the decoder.}
	\label{fig_arch} 
\end{figure}

As depicted in Fig. \ref{fig_arch}, CED-Net is utilized for the estimation of Compton-maps, thereby enabling the estimation of cross scatter.
The CED-Net primarily utilizes a sequence of modules including the Frequency Attetion Module, the Encoder, and Multi-Layer Perceptron (MLP) to achieve the features of total signal
$
z_j = \mathcal{C}(T_j).
$

A notable aspect of CED-Net is the leveraging of the similarity in the distribution of large-angle Compton scatter of dual-ring stationary CT.
The output decoder has been modified accordingly—utilizing the reference Compton-map (for instance, the analytical large-angle Compton scatter signal of a water sphere under the same CT geometry) and spare detector data as input, encoding this information, and concatenating features of varying dimensions into the decoder. This modified structure is termed the Guidance Mapping Module.
Therefore, the predicted Compton-map $\hat{S}^{c}_{j}$ can be expressed as:
\begin{equation}
\hat{S}^{c}_{j}  = \mathcal{R}(z_j,S_0,T^{s}),
\end{equation}
where $\mathcal{R}$ represents Guidance Mapping decoder, $S_0$ represents the reference Compton-map, which is the same for different input, and $T^{s}$ represents the spare detector data.

Unlike traditional serialized decoders, the introduction of spare detector data and reference images into the decoding stage creates a structure named "Guidance Mapping," resembling the traditional Unet architecture. Features extracted from the spare detector data and reference map at various levels are concatenated to the corresponding layers in the decoder.
This approach is based on the fact that the input images for mapping are also scatter images with very similar feature distributions. This method makes full use of the features already present in the spare detector data and reference map.

\begin{figure}[htb]
	\centering
	\includegraphics[width=8cm]{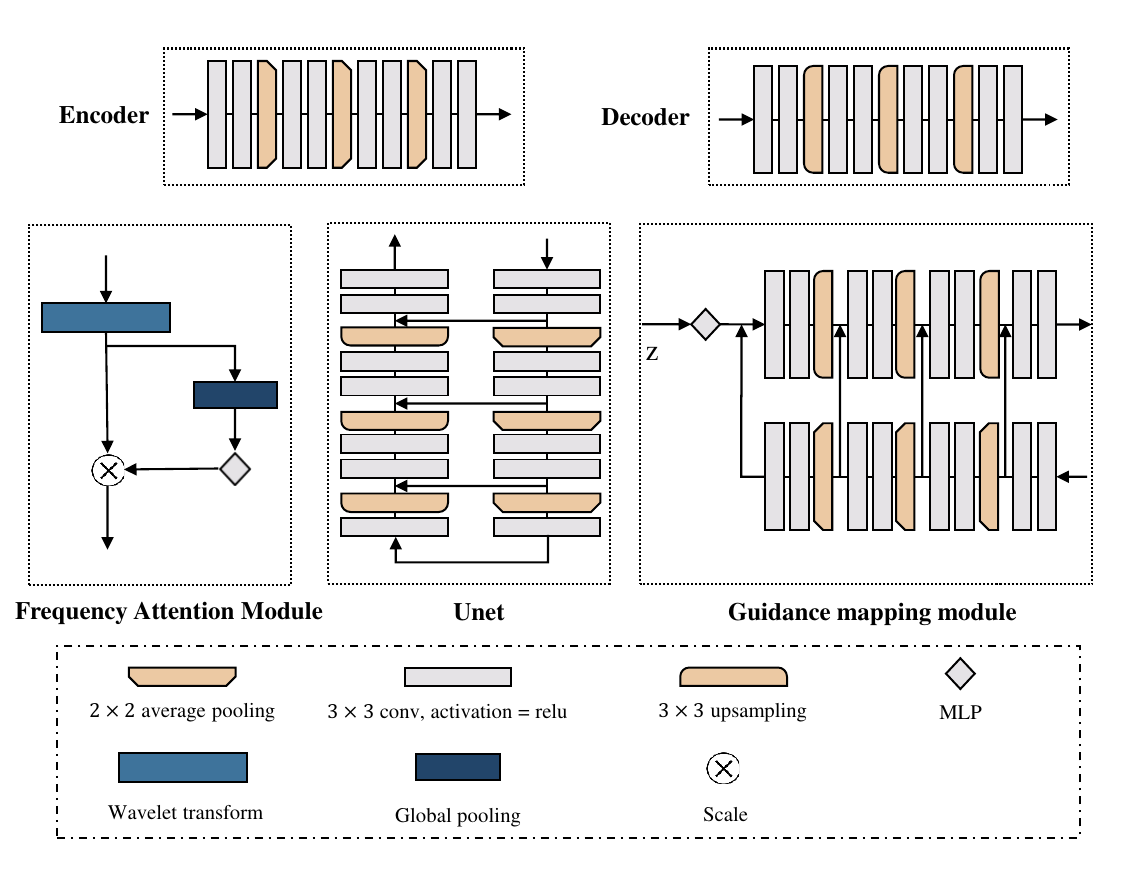}
	\caption{Modules in ComptoNet. Encoder, Decoder, Unet and Frequency Attention Module are common setups. Guidance Mapping Module is similar to the traditional Unet architecture. Features are extracted from the spare detector data and reference Compton-map at different levels to the corresponding layers in the decoder.}
	\label{fig_layers} 
\end{figure}

\subsubsection{Frequency Unet}
In Freq-Unet, frequency Attention Module with frequency attention mechanism and Unet are used for estimate smooth, low-frequency forward scatter signal.
Subpixel Convolution Module\cite{shi2016real} realizes the pixel rearrangement and ensures the output size consistent with the input signal.
The final predicted forward scatter signal $\hat{S}_{j}$ can be expressed as:
\begin{equation}
\hat{S}_{j} 
= \mathcal{F} \left( T_{j} - \sum_{k \neq j} \mathop{sub}\limits_{j} (\hat{S}^{c}_{k}) \right),
\end{equation}
where $\mathcal{F}$ represents Freq-Unet and $\mathop{sub}\limits_{j}(\hat{S}^{c}_{k})$ represents the predicted cross-scatter emitted from the $k$ x-ray source in the $j$ FOV, which is a part of the Compton-map $\hat{S}^{c}_{k}$.

In the literature \cite{maier2018deep,maier2019real,erath2021deep}, a common approach has been to apply average pooling directly to the input signals to retain low-frequency information.
The size of the average pooling is often elusive, and high-frequency information is often directly discarded, which may lead to the network losing too much high-frequency information.
Therefore, how to ensure that the network has a tendency to predict low frequencies without down-sampling the main signal is a very important issue.

Frequency Attetion Module is designed to distribute different frequency components into separate channels\cite{qin2021fcanet}. 
As illustrated in Fig. \ref{fig_layers}, the module first performs a wavelet transform, similar to the Fourier transform, which is capable of extracting different frequency components from a signal.
After the wavelet transform, the corresponding weights for each channel are calculated using global pooling and fully connected layers. Subsequently, these weights are scaled to produce the final output.

\subsubsection{The whole framework}
Combining the above, the final predicted primary signal can be expressed as:
\begin{equation}
	\begin{split}
\hat{P}_{j} 
& = {T}_{j} - \sum_{k \neq j} \mathop{sub}\limits_{j} (\hat{S}^{c}_{k}) - \hat{S}_{j} \\
& = {T}_{j} - \sum_{k \neq j} \mathop{sub}\limits_{j} (\hat{S}^{c}_{k}) - \mathcal{F} \left( T_{j} - \sum_{k \neq j} \mathop{sub}\limits_{j} (\hat{S}^{c}_{k}) \right),
	\end{split}
\label{com_eq}
\end{equation}
where
$$
\hat{S}^{c}_{k}  = \mathcal{R}(\mathcal{C}(T_k),S_0,T^{s}).
$$

(\ref{com_eq}) can be solved by two steps:
\begin{equation}
\begin{split}
\mathcal{R}^{+},\mathcal{C}^{+} 
& = \arg\min_{\mathcal{R}, \mathcal{C}} \mathcal{L} \left[
S^{c}_{j}, \mathcal{R}(\mathcal{C}(T_j),S_0,T^{s}) \right] \\
\mathcal{F}^{+}
& = \arg\min_{\mathcal{F}} \mathcal{L} \left[
S_{j}, \mathcal{F}(P_{j} + S_{j}) \right],
\end{split}
\label{com_eq1}
\end{equation}

\begin{equation}
\begin{split}
& \mathcal{R}^{*},\mathcal{C}^{*},\mathcal{F}^{*} = \arg\min_{\mathcal{R}, \mathcal{C},\mathcal{F}} \\
&  \mathcal{L} \left[
P_{j}, 
{T}_{j} - \sum_{k \neq j} \mathop{sub}\limits_{j} (\hat{S}^{c}_{k}) - \mathcal{F} \left( T_{j} - \sum_{k \neq j} \mathop{sub}\limits_{j} (\hat{S}^{c}_{k}) \right)
\right],
\end{split}
\label{com_eq2}
\end{equation}
where $\mathcal{L}$ is the pixel-wised loss.
$\mathcal{R}^{+},\mathcal{C}^{+},\mathcal{F}^{+}$ are the local optima and $\mathcal{R}^{*},\mathcal{C}^{*},\mathcal{F}^{*}$ are the global optima for the multi-source estimation.

(\ref{com_eq1}) represents the first two cross and forward scatter estimation sub-optimizations and (\ref{com_eq2}) is the fine-tuning for the whole framework.
The multi-source deep scatter estimation framework decouples the cross-scatter and forward scatter, leveraging the consistency between multiple sources to share parameter contributions for the same task. This not only reduces the interdependence between different sub-optimizations but also avoids the complexity of deep learning networks.

\subsubsection{Primary Parameters of Modules}
As shown in Fig. \ref{fig_layers}, the Encoder consists of four stages and utilizes two 3×3 convolutional layers to extract features from the input data, followed by downsampling operations. These operations reduce the spatial dimensions while increasing the depth of the network. Within each Decoder, two 3×3 convolutional layers are employed to extract low-dimensional features from the high-dimensional feature space of the input, followed by upsampling operations. These upsampling operations increase the image scale while reducing the network's depth.

The MLP (Multilayer Perceptron) depicted in Fig. \ref{fig_arch} comprises an input layer, hidden layer, and output layer. Its activation function uses leaky ReLU, which helps prevent the explosion of gradients during training. In Fig. \ref{fig_arch}, the dimension of the feature layer $z$ is 20, achieving a significant dimensionality reduction. Features extracted from the other four FOVs also pass through an MLP, mapping the concatenated 80-dimensional features to 20 dimensions. These features are then dotted with the 20-dimensional features from the main view, thereby obtaining the final intermediate feature layer $z$.

In the Encoder, Decoder, Guidance Mapping, and Unet, convolutional layers for features at different levels are designed with an increasing number of channels from low to high, thereby expanding the depth. In ComptoNet, the modules typically include four stages, with the number of convolutional channels increasing sequentially from $C$, $2C$, $4C$ and $8C$.
For instance, the encoding of CED-Net can be denoted as $\mathcal{C} = \mathcal{C}_{\alpha = 2,\beta = (4,16)}$, indicating that two wavelet transforms are performed in the Frequency Attention Module, and the Encoder module has four stages (consistent with Fig. \ref{fig_layers}) with channel numbers increasing to $16$, $32$, $64$, and $128$ respectively.

\section{Experiments and Results}

\subsection{Monte Carlo simulations}

In order to acquire total raw signal and the counterpart scatter signal in quantity, datasets are generated using a Monte Carlo simulation software package called MCGPU\cite{badal2009accelerating}. The simulation geometry of MCGPU is the common CBCT using a flat panel detector, so geometry changes of MCGPU are made for simulating the geometry of the dual-ring stationary CT. The geometry parameters of the dual-ring stationary CT Monte Carlo simulation are listed in Table \ref{tab:simulations}.

\begin{table}[h]
	\caption{Parameters of Monte Carlo simulations.}
	\begin{center}
		\begin{threeparttable}
			\begin{tabular} {cc}
				\hline
				Parameters & Values \\
				\hline
				Radius of X-ray source ring ($R_1$)   & 110 cm \\
				Radius of detector ring ($R_2$)   & 40 cm \\
				Height of detector ring & 25.6 cm \\
				Z-shift distance of detector ring ($z_1$) & 10.5 cm \\
				Z-shift distance of X-ray source ($z_2$) & -24.7 cm \\
				Detector element size & 1.39 mm $\times$ 1 mm \\
				Size of phantom  & 256$\times$256$\times$128\\
				Voxel size of phantom & 1 mm $\times$ 1 mm $\times$ 1 mm\\
				Photon number of one projection & 2e10\\
				\hline
			\end{tabular}
		\end{threeparttable}
	\end{center}
	\label{tab:simulations}
\end{table}

The datasets were generated using the head, thorax, and abdomen phantoms, which were derived from available CT scans obtained from The Cancer Imaging Archive (TCIA) \cite{clark2013cancer}. The conversion from CT scans to the MCGPU voxel phantom format was performed according to the MCGPU documentation. The simulation utilized a total of $100$ phantoms, comprising $30$ thorax phantoms, $40$ head phantoms, and $30$ abdomen phantoms.
$60$ projections covering $360$ degrees were simulated, resulting in $1200$ scatter-contaminated ring-detector images and their corresponding $1200$ scatter ring-detector images. For each CT phantom, projections were simulated individually.
Five equi-spaced stacks of signals are combined to simulate CT imaging with five X-ray sources operating simultaneously.
The original signal data and scattered signal data output by MCGPU are both measured in electron volts per square centimeter per photon ($eV/({cm}^2 \cdot ph)$). The data was scaled by a factor of $0.4$ to facilitate subsequent deep processing.

\subsection{Experimental Setup}

The simulation totally uses 100 phantoms. Each type of phantoms has two phantoms used as test data. $20$ phantoms (6 thorax phantoms, 8 head phantoms and 6 abdomen phantoms) are using for validation, with the other $74$ phantoms for training. 
The input total signal for scatter estimation is of size $250 \times 256$ (50-deg-wide). CT images are simulated with five X-ray sources emitting simultaneously, resulting in a 22-deg-wide spare detector between the fields of view. The size of the spare detector data is $110 \times 256 \times 5$.
By dividing the framework into two sub-problems, the two networks can be trained separately initially and then concatenated to train the entire framework as a whole, leading to improved predictions of total scatter signals.

For deep cross scatter estimation, given the simpler data structure of cross-scatter, Mean Squared Error (MSE) is chosen as the loss function.
CED-Net was trained for 100 epochs using the Adam optimizer (2e-4) first and then 200 epochs using the Adam optimizer (2e-5).
For deep forward scatter estimation, the target scatter signals are noisy and vary greatly, so Mean Absolute Percentage Error (MAPE) loss function is usually used for deep forward scatter estimation. The network was trained for 200 epochs using the Adam optimizer (2e-4) first and then 200 epochs using the Adam optimizer (2e-5). After FSE and CSE training, the whole framework was trained for 200 epochs using the Adam optimizer (2e-4).
It has been confirmed that all networks have achieved good convergence on the validation set after undergoing the aforementioned training steps. Although the loss function on the training set may still be decreasing, this is attributed more to overfitting.

Upon completion of training, the network's performance is evaluated on the test set. To assess the performance of the proposed deep scatter estimation method, MAPE is used as the evaluation metric.
In our experiment, parameters of ComptoNet are
$$
\mathcal{C} = \mathcal{C}_{\alpha = 2,\beta = (4,16)},
\mathcal{R} = \mathcal{R}_{\beta = (4,16)},
\mathcal{F} = \mathcal{F}_{\alpha = 3,\beta = (4,16)}.
$$
To validate the scatter estimation results in CT images, reconstructions are performed both with and without scatter correction. The projection parameter for reconstruction involves 360 projections with 3600 pixels.
For projections of test CT phantoms, $180$ projections covering $360$ deg are simulated, resulting in $216$ scatter-contaminated ring-detector images, and the corresponding $216$ scatter ring-detector images.
The scatter data ($1800 \times 180$) are interpolated into $3600\times360$ for reconstruction ($512 \times 512$). The numerical calculation primary signal is used instead of the MC primary signals for reconstruction.

\subsection{Results of Cross Scatter Estimation}
\begin{figure}[htb]
	\centering
	\includegraphics[width = 9cm]{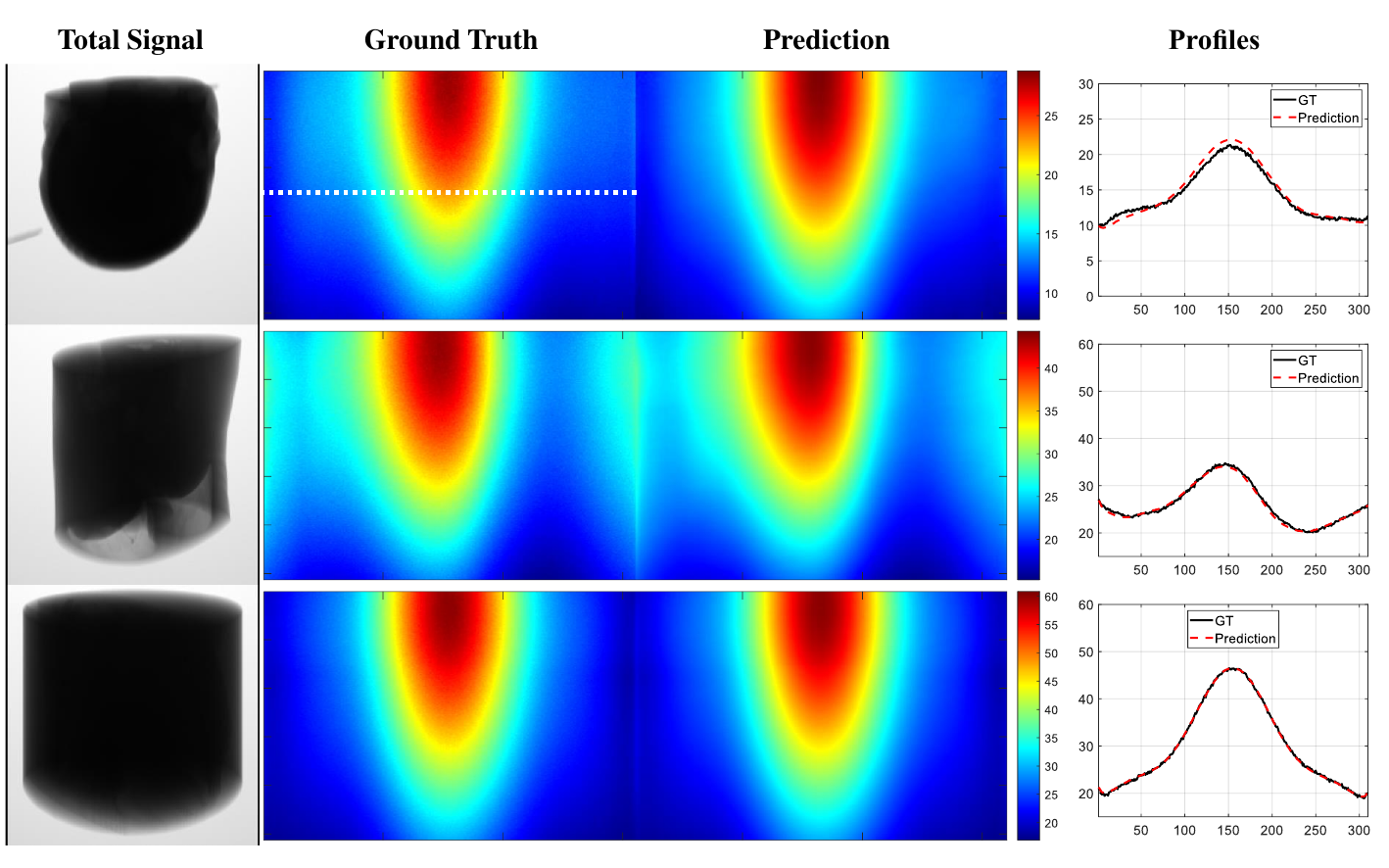}
	\caption{Results of large-angle Compton scatter estimation using CED-Net. The MAPEs of three predictions in the figure are $3.77\%$, $1.37\%$ and $0.65\%$, respectively.}
	\label{fig_CSEresult} 
\end{figure}
As shown in Fig. \ref{fig_CSEresult},  the predicted results show a high degree of consistency with the ground truth. The average MAPE of all test phantoms is $2.84\%$.

\begin{table}[h]
	\caption{Ablation Results of Cross Scatter Estimation using CED-Net.}
	\begin{center}
		\begin{threeparttable}
			\begin{tabular} {cccc}
				\hline
				\multicolumn{2}{c}{Settings} & \multicolumn{2}{c}{Metrics}\\
				Spare Detector Data & Reference Compton-map &MSE&MAPE(\%)\\
				\hline
				\Checkmark & \Checkmark & 0.83 & 2.84 \\
				\hline
				\XSolidBrush & \XSolidBrush & 1.79 & 3.92 \\
				\XSolidBrush & \Checkmark & 0.92 & 2.99 \\
				\Checkmark & \XSolidBrush & 0.92 & 2.81 \\
				\hline
			\end{tabular}
		\end{threeparttable}
	\end{center}
	\label{tab:ablation}
\end{table}

Furthermore, the effectiveness of CED-net is primarily demonstrated through ablation experiments.
The tests are conducted where Spare Detector Data and Reference Compton-map are incrementally excluded from the architecture. Simultaneously excluding Spare Detector Data and Reference Compton-map from the architecture indicates that the Guidance Mapping Decoder approach adopted has reverted to a conventional decoder.

As listed in Table \ref{tab:ablation}, the removal of Guidance Mapping Decoder led to an increase in Mean Squared Error (MSE) from 0.83 to 1.79, which revealed that Guidance Mapping Decoder architecture is crucial for improving the performance of network.
Interestingly, excluding spare detector data or the reference Compton-map results in only a minor decrease in performance metrics, with the MSE ranging from 0.83 to 0.92.
It suggested that there is an overlap in the scattering information provided by Spare Detector Data and Reference Compton-map.
Our ablation experiments underscore the significance of Guidance Mapping Decoder in achieving optimal results and guide future model simplification and optimization efforts.

\subsection{Results of Forward Scatter Estimation}
The effectiveness of our forward scatter estimation network is primarily demonstrated through comparative experiments.
Parameters of Frequency Unet are $\mathcal{F} = \mathcal{F}_{\alpha = 3,\beta = (4,16)}$. The number of parameters of our method is about 1 million. In comparison, Unet(4,16) network has 1 million parameters, while Unet(5,32) network has about 10 million parameters.
Additionally, a network named A3-Unet(4,16) is employed , which, compared to Unet(4,16) network, features an extra average pooling layer with a size of (8,8) at the input. This network can achieve the same image size as F3-Unet(4,16) for the Unet component. However, their channel numbers are different. The number of input channels for F3-Unet(4,16) is $64$, whereas for A3-Unet(4,16), it is $1$.

\begin{figure}[htb]
	\centering
	\includegraphics[width=7.5cm]{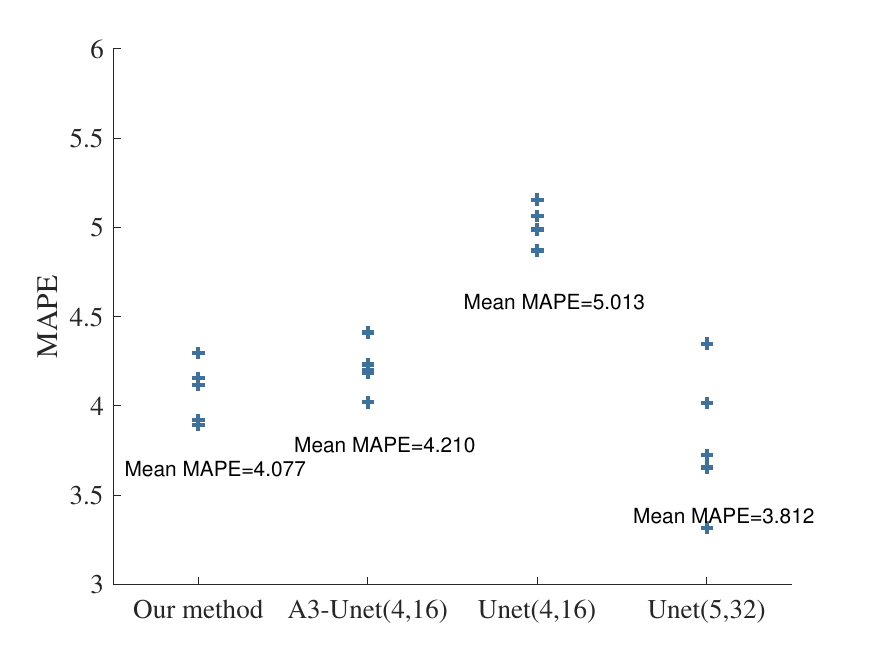}
	\caption{Results of Forward Scatter Estimation using different methods. The parameter of our method is $\mathcal{F} = \mathcal{F}_{\alpha = 3,\beta = (4,16)}$, and A3-Unet(4,16) means that the size of average pooling is (8,8). The number of parameters of our method, A3-Unet(4,16) or Unet(4,16) is about 1 million, while Unet(5,32) network has about 10 million parameters.}
	\label{fig_FSEmape} 
\end{figure}

\begin{figure}[htb]
	\centering
	\includegraphics[width=9cm]{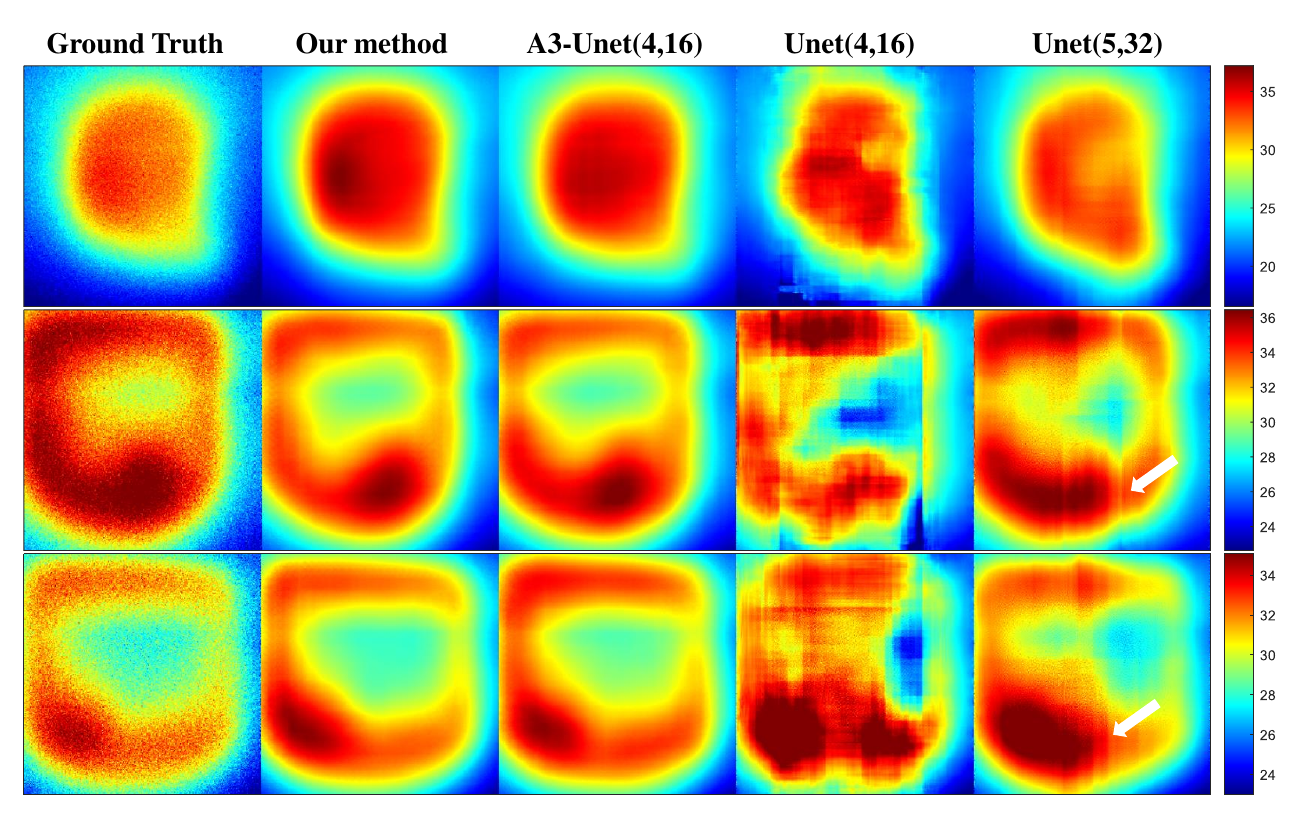}
	\caption{Results of Forward Scatter Estimation using different methods. Our method and the A3-Unet(4,16) prediction results both exhibit high accuracy. The prediction results of Unet(4,16) do not form a coherent shape, and for Unet(5,32), high-frequency artifacts can be observed at the location indicated by the white arrow.}
	\label{fig_FSEresult} 
\end{figure}

As shown in Fig. \ref{fig_FSEmape}, compared to Unet(4,16) and A3-Unet(4,16), our method has achieved better results (mean MAPE = 4.077) in multiple experiments. Although the learning capability of Unet and improve its performance can be enhanced by increasing the number of channels and layers Unet(5,32), whose MAPE could be decreased to 3.812, one can observe that there is a significant variance in the multi-learning outcomes of the network.
In the analysis presented in Fig. \ref{fig_FSEresult}, our approach demonstrates superior and robust performance across various phantoms. The scatter estimations produced by the Unet(4,16) model manifest as more disordered and erratic, whereas the Unet(5,32) model's predictions are marred by high-frequency artifacts.

\begin{figure}[htb]
	\centering
	\includegraphics[width=8cm]{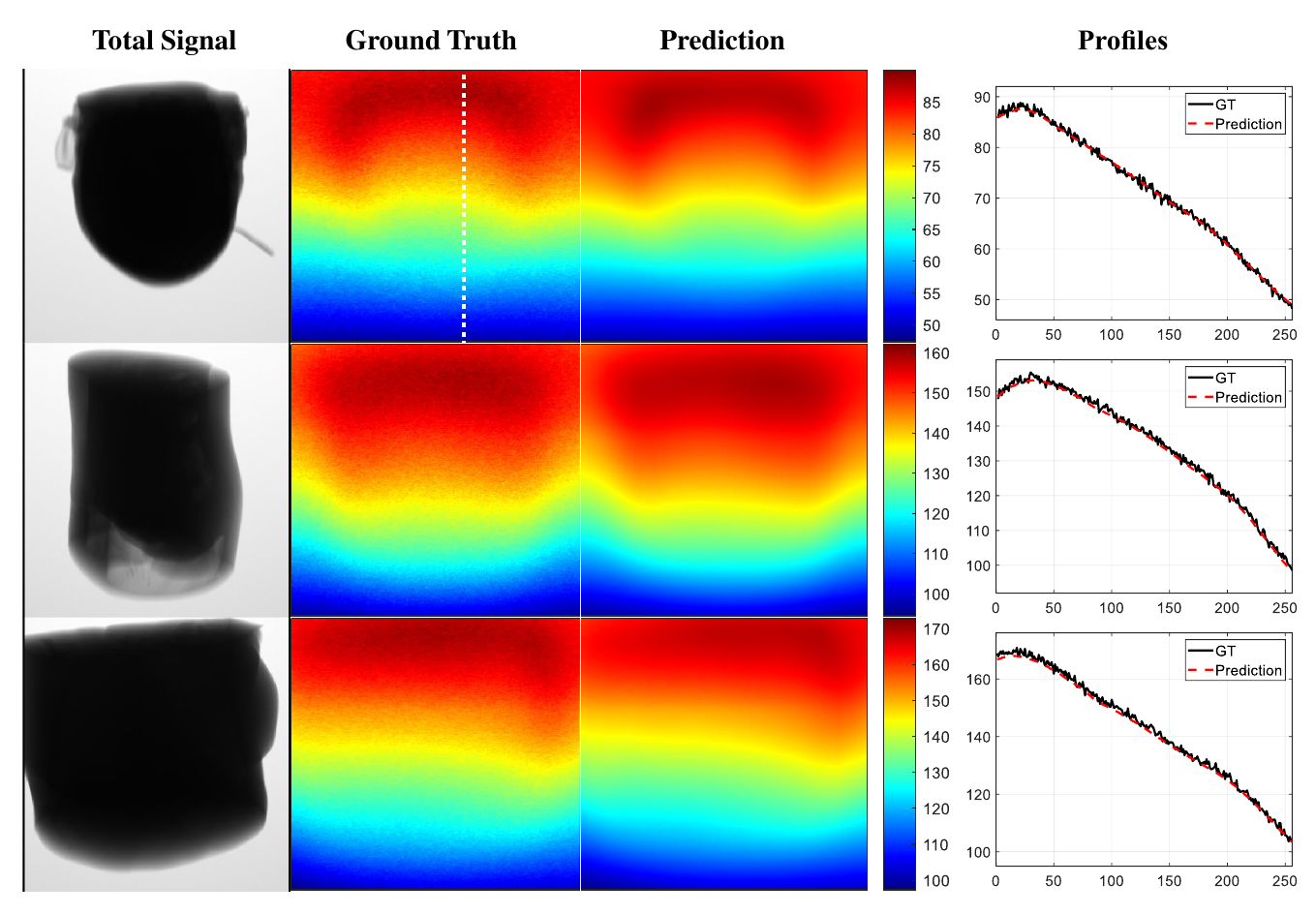}
	\caption{Results of the whole framework. For different phantoms, the predicted results show high accuracy. The MAPEs of three predictions in the figure are $0.93\%$, $0.66\%$ and $0.90\%$, respectively.}
	\label{fig_TSEresult} 
\end{figure}

\begin{figure*}
	\centering
	\includegraphics[width=18cm]{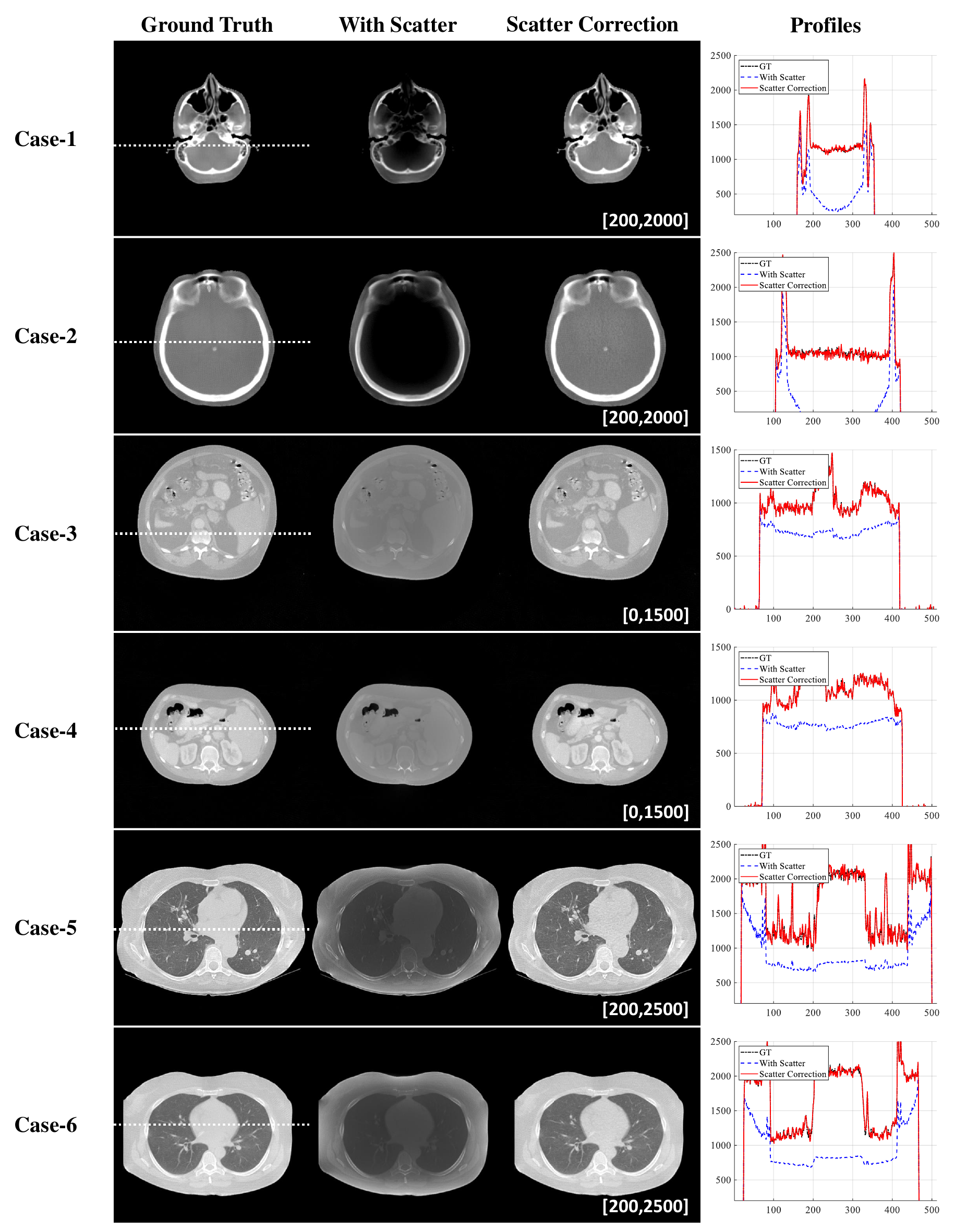}
	\caption{Reconstructions of the deep scatter correction with the proposed framework. The images demonstrate a significant reduction in scatter-induced bias and artifacts, resulting in accurate reconstructions. MSE for reconstructions decrease from $7.6 \times 10^4, 2.2 \times 10^5, 3.2 \times 10^4, 2.9 \times 10^4, 3.3 \times 10^5, 2.9 \times 10^5$ to $25, 102, 17, 12, 2358, 150$.}
	\label{fig_TSErecon} 
\end{figure*}

\subsection{Final Results of ComptoNet}
After pre-training in cross and forward scatter estimation, fine-tuning the entire end-to-end framework as a whole allows us to achieve prediction of the overall scatter signal.
As shown in Fig. \ref{fig_TSEresult}, our framework's predictions demonstrate high accuracy.
\begin{table}[h]
	\caption{The prediction results of ComptoNet for test phantoms under different numbers of photons.}
	\begin{center}
		\begin{threeparttable}
			\begin{tabular}{ccc}
				\hline
				Photon numbers &MAPE(\%)&MAE\\
				\hline
				1e10 & 1.2563 & 1.2989 \\
				2e10 & 1.0239 & 1.0466 \\
				5e10 & 0.8496 & 0.8566 \\
				8e10 & 0.8425 & 0.8488 \\
				\hline
			\end{tabular}
		\end{threeparttable}
	\end{center}
	\label{tab:tse}
\end{table}

Furthermore, the overall performance of ComptoNet framework for scatter estimation in dual-ring MSS-CT is evaluated under different numbers of photons per projection.
As shown in Table \ref{tab:tse}, the framework demonstrates excellent performance under various photon numbers. This indicates that the framework effectively captures and corrects scatter artifacts in the projection data, leading to accurate primary signal estimation.
Simultaneously, as the number of photons increases, it can be observed that the prediction error is minimal, which demonstrates the robustness of our network. The predictions of our network do not become noisy due to the presence of noise in the input. Therefore, as the number of photons increases and the noise in the label data decreases, the performance metrics of our network's predictions also improve (both MAPE and MAE are decreasing).

\subsection{Reconstruction Results}
To mitigate the impact of scatter noise on the final reconstruction result, the number of photons per projection is set to 8e10 in our reconstruction process. The reconstruction algorithm utilized is a three-dimensional weighted approximate FDK algorithm \cite{xia2023generalized}.

Fig. \ref{fig_TSErecon} shows the reconstructed images of test phantoms with and without scatter correction. The images demonstrate a significant reduction in scatter-induced bias and artifacts, resulting in clearer and more accurate reconstructions.
Average MSE for reconstructions decrease from $16.3 \times 10^{4}$ to $444$, representing a reduction of $99.7\%$.
In terms of scatter correction, our method has achieved very accurate results.

\section{Discussions and Conclusions}
This study proposes ComptoNet, an advanced deep learning method specifically engineered for scatter estimation in MSS-CT systems. Leveraging the distinctive distribution of cross scatter, the concept of Compton-map is given and reference Compton-map is employed for realizing a physics-driven deep estimation network.

ComptoNet consists of two networks sequentially: a Conditional Encoder-Decoder Network (CED-Net) for cross scatter estimation and a frequency Unet for forward scatter estimation.
In the deep cross scatter estimation process, reference Compton-map and/or spare detector data are employed for realizing a guidance mapping decoder.
Frequency attention module used in the deep forward scatter estimation preserves the high-frequency scatter information, contrary to conventional down-sampling methods, which leads to improved predictions.
Additionally, the unique spare detector data in dual-ring MSS-CT can also be input into the network, helping ComptoNet achieve a more optimized prediction.

Experimental results substantiate ComptoNet's effectiveness.
For deep cross scatter estimation, the removal of Guidance Mapping Decoder led to an increase in Mean Squared Error (MSE) from 0.83 to 1.79, confirming the significance of Guidance Mapping Decoder architecture.
Excluding spare detector data or the reference Compton-map results in only a minor decrease in performance metrics, demonstrating that there is an overlap in the scattering information provided by Spare Detector Data and Reference Compton-map.
For forward scatter estimation, comparative experiments with other networks, such as Unets, demonstrate our approaches with Frequency Attention Modules have superior and robust performance across various phantoms.
Reconstruction validation further verifies ComptoNet's practical utility, as it effectively mitigates scatter artifacts and elevates the quality of CT reconstructions with average MSE decreasing from $16.3 \times 10^{4}$ to $444$.
The framework's consistent prediction performance across varying photon numbers also demonstrates its robustness and precision.

ComptoNet's end-to-end design eliminates manual feature extraction, streamlining the workflow and enhancing efficiency. Its capability to utilize reference Compton-map and/or spare detector data effectively boosts accuracy and efficiency, underscoring its potential for scatter correction of MSS-CT. 
Such a framework decouples the cross and forward scatter, by leveraging the consistency between multiple sources to share parameter contributions for the same task, result in reduction network complexity.

ComptoNet has demonstrated consistent performance for multi-source scatter estimation. However, there is still room for further improvement. Future research could explore alternative training strategies to boost performance. Expanding the framework to include additional data sources or prior knowledge could increase its robustness and adaptability. Conducting experimental validations in real-world settings is essential to confirm the practicality and effectiveness of these improvements. Lastly, adapting ComptoNet to other multi-source CT architectures and imaging modalities could broaden its impact further.

\bibliography{Ref}
\bibliographystyle{IEEEtran}

\end{document}